# An Improved Deep Convolutional Neural Network by Using Hybrid Optimization Algorithms to Detect and Classify Brain Tumor Using Augmented MRI Images

Shko M. Qader[1,3], Bryar A. Hassan[2], Tarik A. Rashid[4]

[1]Information Technology Department, University College of Goizha, Sulaimani 46001, Iraq

[3]Department of Information Technology, Computer Science Institute, Sulaimani Polytechnic University, Sulaimani 46001, Iraq

[2]Department of Information Technology, Kurdistan Institution for Strategic Studies and Scientific Research, Sulaimani, Iraq

[4]Computer Science and Engineering Department, University of Kurdistan Hewler, Erbil, Iraq

**Abstract**

Automated brain tumor detection is becoming a highly considerable medical diagnosis research. In recent medical diagnoses, detection and classification are highly considered to employ machine learning and deep learning techniques. Nevertheless, the accuracy and performance of current models need to be improved for suitable treatments. In this paper, an improvement in deep convolutional learning is ensured by adopting enhanced optimization algorithms, Thus, Deep Convolutional Neural Network (DCNN) based on improved Harris Hawks Optimization (HHO), called G-HHO has been considered. This hybridization features Grey Wolf Optimization (GWO) and HHO to give better results, limiting the convergence rate and enhancing performance. Moreover, Otsu thresholding is adopted to segment the tumor portion that emphasizes brain tumor detection. Experimental studies are conducted to validate the performance of the suggested method on a total number of 2073 augmented MRI images. The technique's performance was ensured by comparing it with the nine existing algorithms on huge augmented MRI images in terms of accuracy, precision, recall, f-measure, execution time, and memory usage. The performance comparison shows that the DCNN-G-HHO is much more successful than existing methods, especially on a scoring accuracy of 97%. Additionally, the statistical performance analysis indicates that the suggested approach is faster and utilizes less memory at identifying and categorizing brain tumor cancers on the MR images. The implementation of this validation is conducted on the Python platform. The relevant codes for the proposed approach are available at: https://github.com/bryarahassan/DCNN-G-HHO.

## 1. Introduction

Brain tumors are considered a dangerous and scrupulous disease. Tumors can be caused by the abandoned growth of the cancerous cells because they have a critical structural part containing 50 – 100 trillion neurons. Therefore, diagnosing brain diseases becomes complicated because of the presence of the skull around the brain. The stages of the risk involved in a brain tumor depend on the following factors: the size of the tumor, style of the tumor, behavior, position, and status of growth level [1]. Brain tumors are principally classified into two types that are malignant and benign [2]. The tumor, which does not have cancerous cells, is less injurious to humans is known as benign. The tumor contains cancerous cells, which are more harmful to humans than malignant [3].

The process of diagnosing the brain tumor becomes tedious. Some imaging processes regularly do the diagnosing process, these are; single-photon Emission Computerized Tomography (SPECT), Positron Emission Tomography (PET), Computer Tomography scan (CT scan), Magnetic Resonance Imaging (MRI), magnetic resonance spectroscopy (MRS), and Functional Magnetic Resonance Imaging (Functional MRI scan) test similar to the spinal tab, biopsy and Angiogram [4]. To generate images, CT scans utilize the X-ray. CT scan exposes high ionizing radiation to the patients while scanning [5]. This process will lead to an increase in the risk of brain tumors. MRI provides in-depth clear images and often prescribed tests to detect brain tumors because it is a noninvasive approach [6]. MRI uses radio waves and magnetic fields to provide brain images. MR images offer more precise and complete images than CT scans [7]. Therefore, MR imaging is the preferable screening investigation for diagnosing brain tumors since the segmentation process of brain tumors is useful for identifying the growth rate and organizing the clinical procedures. MR images of the brain involves more robust procedures, and it is not to be harmful that are assessment of tissue processing, imaging, and analysis, metabolism, physiology, and function. MRI outputs are expanding the knowledge in medical research to improve the study of the structure of organisms' normal and diseased parts [8]. Brain segmentation includes segregating the different tumor cells into the effective tumor, solid, edema, and necrosis from the normal brain cells, these are GM, WM, and CSF. GM denotes Grey Matter, WM denotes White Matter, and CSF represents the cerebrospinal fluid [9]. Typically, MRI has noises caused by the operator's performance, equipment, and circumference, which would lead to an inaccurate result. The noises presented in the MR images can be eliminated by the denoising approaches, such as Stationary Wavelet Transform (SWT), sharpening filter, and median filter. To detect the brain tumor, exact segmentation becomes a critical process. The manual segmentation process requires more time to detect the brain tumor, as a result, automated or semi-automated approaches are needed to detect the tumor accurately. Different kinds of machine learning approaches have been utilized for identifying the tumor through the MRI along with the k-means, thresholding, Fuzzy C-means (FCM), level set approaches, and kernel extreme learning machine (KELM) [10, 11]. Even though machine learning techniques have some issues, such as interpretation of results, high error susceptibility [12]. To overcome the issues, deep learning-based approaches are creating more attention. Deep learning methods are preferred for the segmentation process, particularly CNN is best for recognizing patterns [13]. Moreover, these approaches absorb the features in terms of hierarchy while comparing with the other statistical methods, including support vector machine (SVM); these depend on the hand-crafted features [14]. Deep CNN approaches are successfully implemented for analyzing the clinical images with the retrieval, segmentation, and classification [15].

Based on the previous works presented above, the following are the difficulties of brain tumor detection encountered by current methodologies:

- The first disadvantage of the current models is their binary categorization of tumors, creating further ambiguity for the radiologist. Additionally, a lack of data hinders the researcher's inability to get reliable results [16].
- Tumor categorization using brain imaging is a challenging task for two reasons. The first reason is that brain tumors exhibit a great degree of variability in size, severity, and form. The second reason is that tumors come in various pathological varieties, showing the same symptoms [17].
- In [18], a method called Adaptive Convex Region Contour (ACRC) was developed for segmenting brain tumors using magnetic resonance imaging. The SVM was used to categorize the slice to determine whether it was expected or pathological. Furthermore, the technique demonstrated enhanced performance and resolved high dimension, short sample size, and nonlinearity problems. However, this method required tremendous feature values, introducing noise into the classification results.

- In [19], a convolutional neural network-based deep learning model was developed to identify different brain tumors. While deep learning methods improve tumor classification, they need a large amount of training data for analysis. Moreover, the computational cost and training time associated with brain tumor classification are substantial.
- Segmentation and identification of tumorous areas from brain MRIs are complicated and take much time for analysis. In addition, the method's accuracy is significantly impacted when identifying brain tumors [20].

The images used to identify brain tumors are important since brain tumors are considered the most severe illness. Accurate diagnosis of a brain tumor may enable the tumor's precise location in the brain to be determined. As a result, the mortality rate will decrease. Thus, it is critical to detect hidden patterns to improve images and the diagnostic process. However, obtaining an appropriate diagnosis for the different lesion types becomes difficult, regarded as the primary issue. To address this issue, traditional methods are employed, including presenting a DCNN based on the enhanced HHO methodology. This study aims to provide a method for detecting brain tumors that incorporate a DCNN presentation based on the improved HHO. The proposed approach automatically detects brain cancers in magnetic resonance images, with feature extraction doing the classification. To begin, magnetic resonance images of the brain are regarded as input for pre-processing to create images suitable for subsequent processing. The pre-processed images are then subjected to the segmentation process using the Otsu thresholding method. Once the segmentation procedure is complete, extracting the texture and statistical characteristics from each segment could be carried out. The characteristics include the tumor's size, variance, and mean. Moreover, these retrieved features are formulated in the feature vector. Finally, the detection of brain tumors is carried out using DCNN and the feature vector, where the Deep CNN method is trained using improved HHO. The suggested approach enhanced HHO combines the standard GWO and HHO optimizations to benefit practical classifier training.

This research is distinguished from earlier ones by the following significant contributions:

- G-HHO improves DCNN to detect and classify brain tumors from augmented MRI images effectively;
- The G-HHO has the features of GWO and HHO;

- Otsu thresholding is used to segment the tumor part, emphasizing the identification of brain tumors;
- Experiments are conducted by the Python platform on vast augmented MRI image datasets;
- The performance of the suggested DCNN-G-HHO is improved and enabled better detection and classification results.

The remainder of this research paper is ordered as follows: Section 2 describes the traditional brain tumor detection methods used in the literature and the difficulties encountered, which served as the motivation for creating the suggested approach. Section 3 describes the recommended technique for detecting brain tumors using DCNN based on improved HHO. Section 4 proposes G-HHO to improve DCNN to classify brain tumors. Experimental results are also discussed in Section 5. Section 6 compares the suggested approach to alternative ways, and Section 7 concludes remarks of this work.

## 2. Literature review

Brain tumors are a particularly aggressive kind of cancer that may result in significant problems in the human body. Thus, early, and accurate identification of a brain tumor may significantly improve the chance of survival. However, proper identification of various tumor types is a difficult problem. Thus, developing an efficient tumor representation via an optimization algorithm is critical toward successful brain tumor identification. Ten current methods for detecting brain tumors are examined in this research work, and the shortcomings of each approach serve as inspiration for developing a new brain tumor detection strategy. Thaha MM et al. [21] have presented the enhanced convolutional neural network for the computerized segmentation process using the BAT algorithm. Furtherly, they have used skull stripping and image enhancement algorithms for the pre-processing. Chen S et al. [22] have presented the segmentation models, such as multi-level deep medic, dual force training scheme, auxiliary classifier, and multi-layer perceptron-based post-processing approach to improving the standard of the hierarchical features for detecting brain tumors. Nonetheless, they failed to provide enhanced segmentation capability for the deep architecture to avoid the individual post-processing steps. Anaraki AK., Ayati M., and Kazemi F., [23] have presented the hybrid method based on the Genetic Algorithm (GA) and convolutional neural network (CNN) to categorize the various category of glioma by MRI. Furtherly, an ensemble algorithm was used to reduce the prediction error variance. Talo M et al. [24] have presented pre-trained models for automatically classifying MR images, such as AlexNet,

egg-16, ResNet -18, ResNet- 34, and ResNet-50. The images are categorized into normal, cerebrovascular, neoplastic, degenerative, and inflammatory diseases classes. Toğaçar M et al. [25] have presented the convolutional neural network model called BrainMRNet based on the attention modules using the augmentation and hyper column approaches. Where Accessible magnetic resonance images were utilized to identify the brain tumor along with the brainMRNet model. Özyurt F et al. [26] have presented the hybrid method of Super-resolution fuzzy-c-means CNN (SR-FCM-CNN) based on the convolutional neural network, and the learning machine algorithm, and fuzzy c-means to identify the brain tumor. Where features are extracted by the squezeeNet architecture from the CNN model. Sharif M et al. [27] have presented the brain surface extraction method to get skull removed images. They have presented particle swarm optimization (PSO) for segmentation purposes. They have used GA for feature selection. Also, artificial neural networks and classifiers were used to categorize the tumor levels. Navid Ghassemi et al. [28] have presented the deep learning technique to classify brain tumors in MRI images. Where deep neural network was pre-trained in a generative adversarial network (GAN) to separate the features and analyze the structure of MR image within its convolutional layer. Zeineldin RA et al. [29] have presented the Deepseg technique from the genetic deep learning architecture to provide computerized identification and segmentation of tumors through the FLAIR MRI data. For the extraction process, they were used CNN. The CNN models, such as dense convolutional network (DenseNet), NASNet, and residual neural network (ResNet) were used based on the improved architecture of U-Net. Vijh S et al. [30] have presented the particle swarm optimization with the Otsu algorithm for automatically diagnosing the brain tumor. An anisotropic diffusion filter was used to eliminate the noise, and a convolutional neural network was used for the classification process. Finally, the most recent research work has presented an optimization-based approach, called Whale Harris Hawks optimization (WHHO), for brain tumor diagnosis utilizing magnetic resonance imaging (MR) images [20]. Segmentation was carried out utilizing cellular automata and rough set theory, and several parameters were retrieved from the segments. Furthermore, brain tumor diagnosis was accomplished via DCNN, which was trained using the proposed WHHO. The suggested WHHO method was created by combining the Whale Optimization Algorithm (WOA) with the HHO. With a maximum accuracy of 0.816, a maximum specificity of 0.791, and a maximum sensitivity of 0.974, the suggested WHHO-based DeepCNN outperformed previous approaches.

Table (1): Summary of current methods for brain tumor detection

| Reference, author(s) | The algorithm | Concluding remarks | Limitations |
|---|---|---|---|
| [21], Chen S et al. | Dual-force convolutional neural networks | The proposed approach promoted the segmentation performance of BRATS 2017 and BRATS 2015 datasets. | Failed to provide enhanced segmentation capability for the deep architecture. |
| [22], Thaha MM et al. | CNN | CNN results are promising in BRATS 2013 and 2015 datasets in terms of accuracy and computational time. | The CNN components should be further investigated to develop a more robust algorithm for brain tumor segmentation using MRI data. |
| [24], Talo M et al. | CNN | CNN obtained the best classification accuracy among the five pre-trained models. | The fewer available annotated data. |
| [23], Anaraki AK, and Ayati M, Kazemi F | CNN based on GA | The findings demonstrated the suggested method's efficacy in classifying brain tumors using MRI images. | The proposed algorithm should be applied on larger datasets with several tumor types and other CNN structures and deep learning algorithms to improve performance. |
| [25], Fatih Özyurt et al. | BrainMRNet | BrainMRNet was more successful than the current CNN models (AlexNet, VGG-16, and GoogleNet). | BrainMRNet should be used on different medical images and in various fields. |
| [26], Özyurt F et al. | SR-FCM-CNN) | SR-FCM-CNN determined that brain tumors have been better segmented and removed using the SR-FCM method. | The performance of SR-FCM-CNN varies depending on the training dataset. |
| [27], Sharif M et al. | Ensemble algorithm | The suggested approach performed better in comparison with existing methods. | The ensembled algorithm can be extended for the detection of sub-structures of a tumor. |
| [29], Zeineldin RA et al. | DeepSeg | DeepSeg showed successful feasibility and comparative performance for automated | The validation of the DeepSeg framework should be extended on more image datasets from other different MRI modalities. |

| | | brain tumor segmentation in FLAIR MR images. | |
| --- | --- | --- | --- |
| [30], Vijh S | Hybrid OTSU+APSO | The suggested algorithm yielded a 98% accuracy rate, superior to current methodologies. | The model should be further optimized for various data modalities, and additional metaheuristic algorithms will be utilized to enhance the diagnosed system's performance. |
| [20], Rammurthy D, and Mahesh PK | WHHO-based DeepCNN | The WHHO-based DeepCNN outperformed other algorithms with maximal accuracy, specificity, and sensitivity. | Additional brain tumor datasets should be utilized to maximize the suggested method's efficiency. Also, several optimization approaches will be investigated to improve the efficiency of the introduced method. |

Based on the above-presented table, current methods' difficulties are (i) Insufficient precision in identifying brain tumors. (ii) Inaccurate classification of brain tumors. (iii) Lack of using large MRI image datasets as input data. Therefore, this paper provides the following contributions for mitigating the limitations of the current algorithms:

- G-HHO enhances DCNN to identify and classify brain malignancies from enhanced magnetic resonance imaging pictures more efficiently.
- G-HHO provides all the good characteristics of both GWO and HHO.
- Otsu thresholding is used to segment the tumor portion, emphasizing the detection of brain tumors.

Additionally, Table (2) indicates the primary questions of this work and how we have addressed them.

**Table (2): The research questions and mitigations of this work**

| # | Research question | Mitigation |
|---|---|---|
| 1 | Inadequate accuracy in the identification of brain tumor | Increasing sufficient detection of brain tumors. |
| 2 | Inaccurate brain tumor categorization | Increasing accuracy of classifying brain tumors. |
| 3 | A deficiency in the use of big MRI image datasets as input data | Augmented MRI images are used as the input dataset. |

## 3. The proposed methodology

During brain tumor detection, the accuracy level of the images is considered an essential one because brain tumors are considered the most critical disease. Accurate diagnosing of a brain tumor may identify the exact affected portion of the brain. Thus, it will reduce the death rate. Therefore, it is essential to identify the hidden patterns to enhance the quality of images and enhance the diagnosis process. But attaining accurate diagnosis for the various lesion cases becomes a complex one, considered the main problem. To overcome this problem, conventional techniques are used, such as a DCNN based on the improved HHO technique, are presented. The presented model automatically identifies brain tumors from the MR images, where feature extraction executes the classification. Firstly, MR images of the brain are considered input for pre-processing to make the images for further processing. After that, the pre-processed images are induced to the process of segmentation through the Otsu thresholding technique. Once the segmentation process is completed, then the features in every segment as per texture and statistical are extracted. The features include the size of the tumor, variance, and mean. Furthely, in the feature vector, these extracted features are formulated. At last, by DCNN with the feature vector, brain tumor identification is performed where the presented enhanced HHO trains the Deep CNN technique. The proposed technique of improved HHO combines the conventional GWO and HHO to get the advantages of the two optimizations for effectual training the classifier. The overall execution process of the proposed DCNN-G-HHO is shown in Figure (1).

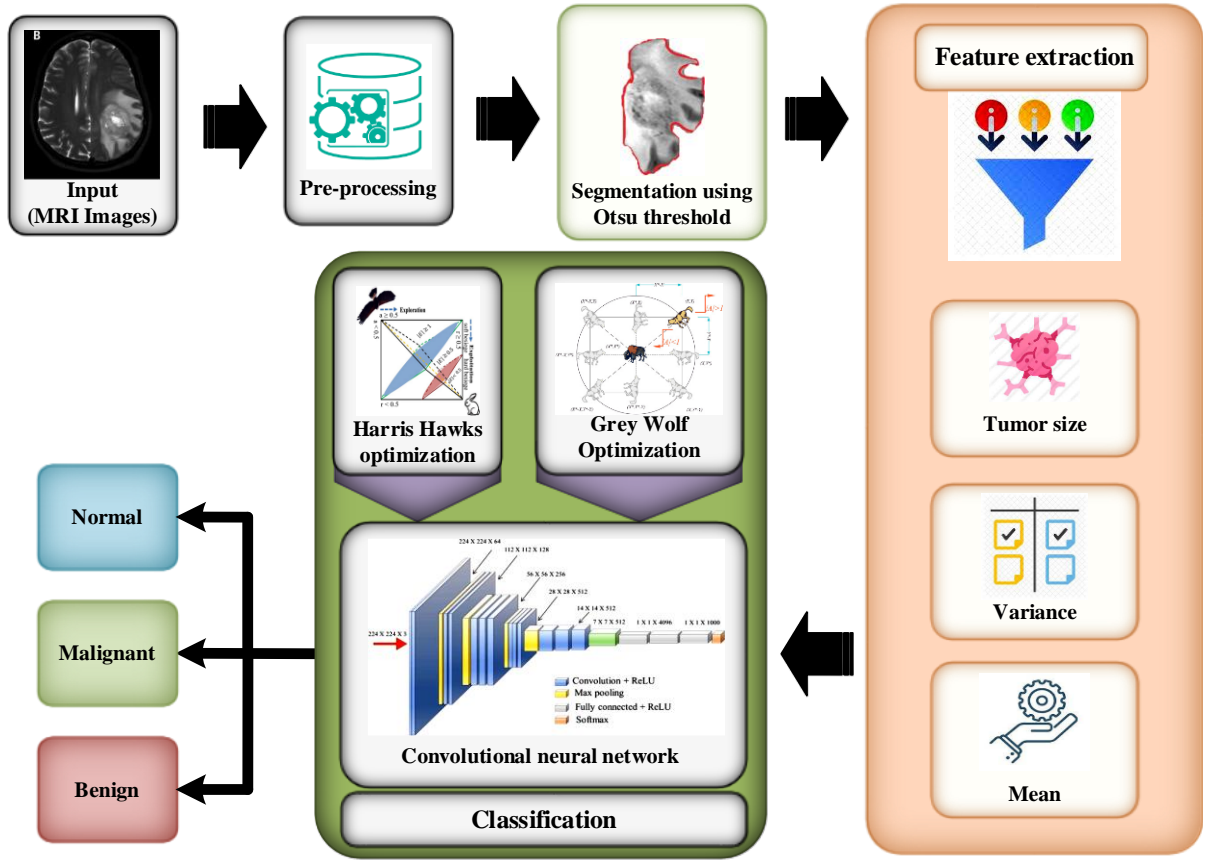

**Figure (1): The proposed DCNN-G-HHO model for brain tumor detection**

Consider the *D* database and the g number of input MRI images, and it can be mathematically represented in Equation (1).

$$D = \{I_1, I_2, \ldots\ldots I_e, \ldots\ldots, I_g\} \tag{1}$$

Here, $I_e$ denotes the $e^{th}$ image and $g$ denote the MRI image's total count in the Database.

In the following sub-sections, three components (Pre-processing, Segmentation using Otsu threshold, and feature extraction) of the proposed DCNN-G-HHO are detailed.

The proposed method consisted of the following stages:

### 3.1. Pre-processing

The primary usage of pre-processing is providing a further smooth process of the input image. The pre-processing step is more required to make the images appropriate for the identification process for processing the image. Furtherly, the pre-processing is executed to remove the artifacts and noises presented in the extracted images. Thus, the pre-processing step performs image enhancement to improve the contrast of the images to detect the brain tumor. Then, for the segmentation process, the pre-processed images are given for fetching the essential features which are more suitable for identifying the tumor.

### 3.2. Segmentation

Using the cellular automata model, the pre-processed images are given to the segmentation module to produce segments. The pre-processed images contain various elements, and every segment denotes the separate regions. In brain tumor identification, Otsu thresholding is employed to discover the cancerous portions by each component.

Otsu algorithm can be distinguished among the foreground and background of the images by setting the image above the calculated value of grayscale. Before executing the segmentation algorithm, the Otsu algorithm needs to calculate the image's grey level histogram. The primary purpose of the Otsu technique is to discover the threshold that reduces the intra-class variance, and it can be described as the two-class weighted sum of variances as shown in Equation (2) [31].

$$\sigma_\omega^2 = \omega_0(t)\sigma_0^2(t) + \omega_1(t)\sigma_1^2(t) \qquad (2)$$

Where, $\omega_0$ and $\omega_1$ represent the weights that are the possibilities of the classes, and by the threshold value *t*, these weights are separated. $\sigma_0^2$ and $\sigma_0^2$ represent the variances of two classes. After that, the images are binarized more than the threshold value, and covering the holes presented in the segmented image can be made by executing the morphological operations. Furtherly the segmented images are masked upon the original images; thus, the segmented tumor is attained.

### 3.3. Feature extraction

By now, numerous research on deep feature extraction approaches has been conducted, with promising results for image classification. One of the feature extraction models was proposed by [32] to combine both approaches into a unified framework and to develop an unsupervised linear feature selective projection (FSP) with low-rank embedding and dual Laplacian regularization exploit the intrinsic relationship between data and suppress the impact of noise. The proposed approach outperformed existing state-of-the-art algorithms for feature extraction in the presence of different types of noise. Another study conducted by [33] for feature extraction, took into consideration both the minimizing of reconstruction error and the maximizing of variance in a unified framework. The experimental findings from multiple databases demonstrated the model's efficacy. However, it was expected to explore a more straightforward but efficient method for its selection. Last but not least, unsupervised multi-view feature selection using cross-view local structure-maintained diversity and consensus representation learning was suggested by [34]. Projecting each original data view onto a shared label space with a consensus component and a diversity part captures shared and distinct knowledge across images. Numerous pre-defined view-specific similarity graphs are employed to train a shared similarity network across views. Despite its shortcomings, the suggested method's effectiveness is shown by a parameter sensitivity analysis on real-world Multiview datasets.

To address the limitation of the above studies, the feature is obtained from the MRI input images to create a feature vector. When attaining the segments, the features are extracted, and it carries out the pre-processed images for every segment. The feature extraction assures the effectual identification of brain tumors because it involves texture and statistical features. For extracting the features, every segment considers it for attaining better accuracy in detecting brain tumors. As earlier discussed, from each segment the feature extraction proceeds in terms of tumor size, mean, variance. This can be described below:

### 3.3.1 Mean

The mean value can be computed by summarizing the entire numbers presented in the dataset and dividing it by the total count of the number presented in the dataset. The mean value is considered the vital parameter for the image segmentation process. A two-layer feed-forward neural network can forecast the growth of the tumor within the 16% MSE. The mean can be computed by calculating the pixel's average presented in the image, and it can be mathematically expressed as formulated in Equation (3).

$$\mu = \frac{1}{|d(S_n)|} \times \sum_{n=1}^{|d(S_n)|} d(S_n) \qquad (3)$$

Here, $n$ represents the total segments, $d(S_n)$ denotes the every segments pixel values and $|d(S_n)|$ represents the total pixel of the segment.

### 3.3.2 Variance

Measuring the spread among the numbers of a dataset is known as a variance. This variability implies measuring the distance between each number from another number in a dataset. Variance can find the reasons for the deviation in the brain, and it helps recognize the tumor cells in the brain. The variance features are computed and mathematically expressed in Equation (4) according to the mean value.

$$\sigma = \frac{\sum_{n=1}^{|d(S_n)|} |S_n - \mu|}{d(S_n)} \tag{4}$$

### 3.3.3 Tumor Size

From the Brain MR images, tumor size extracts the features to detect the brain tumor. Tumor segmentation process from the brain MRI contains various covering pathology such as MRI material science, perception of radiologist and analysis of the image. A brain tumor can be of any size; it has different shapes and may locate on any portion and may show deviation in the image intensities.

The variable Q indicates the tumor affected boundary, the derivation of the area of feature vector from the tumor-dominated portion. Q is mathematically represented in Equation (5).

$$Q = \frac{\pi}{4}(L \times W) \tag{5}$$

Where L represents the length of the tumor size, W denotes the width of the tumor size.

### 3.3.4. Formation of the feature vector

The below expression describes the set of texture and statistical features. Therefore, the extracted features of every segment are represented as shown in Equation (6).

$$J = \{\mu, \sigma, \varepsilon, \mu\} \tag{6}$$

Where J represents the extracted feature vector by each segment such as $\mu, \sigma, \varepsilon, \mu$ denotes the mean. $\sigma$ denotes the variances. $\varepsilon$, indicate the tumor size. After that, the extracted feature vector is fed to the DCNN, which categorizes the images based on the given features, and forms the class label. The classifiers are used to formulate the class label and segregate the cancerous and noncancerous cells concerning the input image.

## 4. Proposed G-HHO to improve DCNN

This section, presents the proposed G-HHO technique to detect the brain tumor where the detection process is executed with the consideration of a feature vector. Moreover, the extracted features are further processed for classification by DCNN trained with enhanced HHO to classify the tumor. In the improved HHO, the parameters of GWO are utilized; therefore, the performance of HHO can be improved. Based on the social behavior of the GWO is developed. The GWO algorithm is more suitable to get optimum global results because it gives a better convergence rate [35]. Moreover, GWO can resolve real-time problems and complicated search spaces. Also, HHO is adopted from Harris's hawks' helping and chasing behavior. This approach is more effective for witnessing optimization problems and providing optimum results. Even though the HHO technique has some issues in the search space, such as multi-modality, deceptive optima, and optimal local solution, the proposed work that combines the GWO with HHO (G-HHO) will give better results it limits the convergences rate and enhancing the performance.

Furthermore, our proposed hybrid approach provides better exploration and exploitation than other meta-heuristic algorithms, such as minimal sequential optimization (SMO), BAT, and PSO. The exploration and exploitation of search areas are calibrated to emphasize exploitation as the iteration counter grows. The suggested DCNN-G-HHO model incorporates a trade-off between the exploitation and exploration phases to identify the optimal solutions and converge to the global optimum. Regardless of the common features among the current meta-heuristic algorithms, The DCNN-G-HHO adds a feature of improved searching spaces. The search processes are divided into exploration (diversification) and exploitation (intensification). The algorithm should maximize the use and promotion of its randomized operators throughout the exploration phase to examine different areas and sides of the feature space thoroughly. Thus, the exploratory behaviors of a well-designed optimizer should be sufficiently random to efficiently distribute more randomly produced solutions around the problem topography during the early stages of the search process [36]. Typically, the exploitation step follows the exploration stage. The optimizer concentrates on the neighborhood of higher-quality solutions situated inside the feature space during this phase. It accelerates the search process inside a specific location rather than throughout the terrain. A well-organized optimizer should strike a sensible and delicate balance between exploration and exploitation inclinations. Otherwise, the likelihood of being caught in local optima and suffering from immature convergence downsides grows.

The algorithmic stages of suggested improved HHO and the structure of the Deep CNN model are defined in Equation (7).

$$T = \{T_1, T_2, \ldots, T_h, \ldots T_l\} \tag{7}$$

Here, l denotes the total count of the convolutional layer of the Deep CNN model and $h^{th}$ a convolutional layer of the Deep CNN can be represented by the $T_h$. Using the units (p,w), the output can be derived and expressed in Equation (8).

$$\left(T_\vartheta^h\right)_{p,w} = \left(v_\vartheta^h\right)_{p,w} + \sum_{p=1}^{W_1^{P-1}} \sum_{z=-r_1^u}^{r_1^u} \sum_{s=-r_2^u}^{r_2^u} \left(X_{f,p}^u\right)_{z,s} * \left(T_v^{h-1}\right)_{p+z,\ w+s} \tag{8}$$

Where the conv operator can be denoted as * and is used for permitting the extraction process of pattern from the results taken out from the adjacent convolutional layer, feature maps can be represented as $\left|\left(T_p^{u-1}\right)_{m+z,r+s}\right|$. total count of feature maps can be denoted by $W_1^{P-1}$ and $\left(X_{f,p}^u\right)_{z,s}$ Denotes the weights. Where the trained by the proposed G-HHO.

The following three sub-sections are given below for the proposed algorithm (G-HHO).

### 4.1. Rectified linear unit and pooling layer

Rectified linear unit and pooling layer (ReLU) is considered an activation function because its contribution is highly significant in simplicity and efficiency. Moreover, ReLu in Deep CNN provides speed processing with large networks. The output obtained through the ReLu layer is given based on providing feature maps in Equation (9).

$$T_f^u = fun\ T_f^{u-1} \tag{9}$$

Where $T_f^u$ represents the input and $T_f^{u-1}$ represents the output and $fun\ ()$ represents the $u^{th}$ layer activation function.

### 4.2. Fully connected layers

To initialize the classification process of objects, the patterns of the convolutional and pooling layers are provided as an input to the completely connected layers. The output of the completely corresponding layer can be mathematically expressed in Equation (10).

$$S_f^u = Z\left(a_f^u\right) \text{ where } a_f^u = \sum_{p=1}^{W_1^{P-1}} \sum_{m=1}^{W_2^{P-1}} \sum_{n=1}^{W_3^{P-1}} \left(v_{f,p,m,n}^u\right) \cdot \left(T_f^{u-1}\right)_{m,n} \tag{10}$$

where, $V_{f,p,m,n}^u$ represents the weights (m,n), which is connected with the $u-1$ layer's $p^{th}$ feature map and u layer's $f^{th}$ unit. By the proposed G-HHO approach, these weights are optimally tuned.

### 4.3. The network architecture of the proposed DCNN model

The proposed work has six architecture layers that incorporate three convolutional layers, two fully connected layers, and the last layer, one classification layer. These convolutional layers are connected by three rectified linear operator (Relu) layers and max-pooling layers. The initial convolutional layer has a total number of fifty-two filters with a size value of $7 \times 7$ pixels, a stride of 2 pixels, and no padding is coming after ReLu. A max-pooling layer has a value of $3 \times 3$ region and 2 pixels for stride. The result obtained through convolutional layer 1 is $52 \times 69 \times 69$; this result is considered an input value for the next layer. The second convolutional layer has filters 256, and the size of them are $5 \times 5$ pixels, two pixels for stride, and similar to the first convolutional layer, no padding is set after Rectified Linear unit as well as ReLu, and $3 \times 3$ regions as a value of max pooling and 2 pixels for stride. The resulting outcome for the second convolutional layer has obtained $256 \times 15 \times 15$ pixels as size as the first convolutional layer. This result is taken as an input to the third convolutional layer. The third convolutional layer is set with 156 filters with a 3 $\times$ 3 region and 2 pixels for strides. The third layer result goes to the fourth layer as input; this layer is the fully connected layer with neurons 512. This layer comes after the dropout layer with a probability value of 0.5. At last, the classification layer shows the results with two classes. Using appropriate techniques to tune the hyperparameters of DCNN is significant to improve the classification technique [37]; therefore, the combination of Grey Wolf Optimization and Harris Hawks Optimization is considered to enhance the performance. The following section illustrates the GWO and HHO in detail.

#### 4.3.1. HHO

HHO technique is considered to enhance the performance of the deep learning technique [38]. HHO is mainly relying on two stages exploration and exploitation. As represented in Figure (2), the following section provides the step-by-step procedure of HHO.

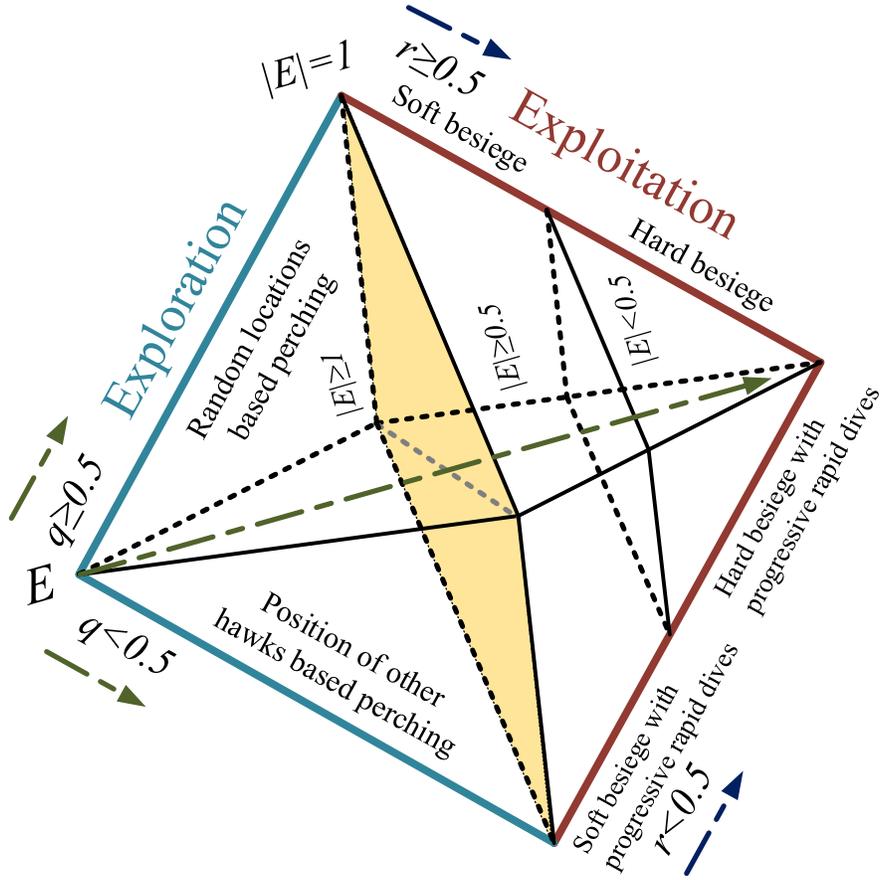

**Figure (2): Various stages of Harris Hawks Optimization (Extracted from** [28]**)**

### 4.3.2. exploration

The first stage is the exploration stage, where the entire Harris Haws represent the candidate solution. Here, fitness value is based on targeted prey that may change every iteration. Equation (11) represents the exploration performance [38].

$$X(i+1) = \begin{cases} X_{rsolu}(i) - r_1|X_{rsolu}(i) - 2r_2 X(i)| & q \geq 0.5 \; rule \quad (i) \\ \left(X_{prey}(i) - X_m(i)\right) - r_3(LB + r_4(UB - LB)) & q < 0.5 \; rule \quad (ii) \end{cases} \quad (11)$$

Where:

$X(t+1)$  => Represents the position of Hawk's in the next iteration i.,

$i$  => Iteration,

$X_{prey}$  => Prey's position,

$X_{rsolu}(i)$  => Random solution selection process based on current population,

| $X(i)$ | => | Represents the position vector of Hawk's based on current iteration, |
| $r_1, r_2, r_3, r_4,$ and $q$ | => | Random scaled factor within range of [0, 1] ( updated in every iteration), |
| $LB$ | => | Lower bound of variables, |
| $UB$ | => | Upper bound of variables, |
| $X_m$ | => | Solution's average number. |

Based on two rules, the position of hawks is developed by this targeting strategy within $(UB - LB)$.

Two strategies are adopted to determine the Hawks' position (UB-LB). The initial rule is Hawks based solutions are generated randomly based on the current population and other hawks. The second rule is a solution prepared based on the Hawk's average position, prey's location, and the random scaling factor. These steps help to enhance the randomness of the rule. The randomly scaled action in terms of length is fed to the Lower Bound (LB). Therefore, various feature spaces can be explored, and the random scaled feature enables the diversification technique. Equation (12) represents the average solution position [38].

$$X_m(i) = \frac{1}{N}\sum_{i=1}^{N} X_i(i) \qquad (12)$$

Where:

| $X_m(i)$ | => | Coordinate average of solutions which is presented in the current iteration, |
| $N$ | => | Whole possible results, |
| $X_i(i)$ | => | Each result's location in $i^{th}$ iteration, |

The Hawk's prey catching process is based on a random solution that can be achieved through Equation (10 (i)). The entire hawks show the best solution can be done through Equation (10 (ii))

### 4.3.3. Transition from exploration to exploitation

Based on the rabbit's energy ($E$), the movement of Harris hawk's optimization from exploration to exploitation is described in this stage.

In this algorithm, the prey (Eng) energy is significant to execute the optimization through the exploration and exploitation shown in the following Equation (13) [38]. The preys escaping pattern will be reduced due to the action of HHO. The energy reduction is indicated from [1, -1] is indicated as $Eng_0$.

$$Eng = 2Eng_0\left(1 - \frac{i}{I}\right), \quad Eng_0 \in [-1,1] \tag{13}$$

Where:

    $I$    =>    Overall iterations,

    $i$    =>    Current iteration.

### 4.3.4. exploitation

At the time of hunting, the prey takes maximum effort to escape from being attacked. Therefore, HHO executes four different techniques to confuse its prey in the exploitation stage, where the position is highly significant. These techniques are soft besiege, soft besiege with progressive rapid dives, hard besiege, and hard besiege with advanced quick dives.

Based on variables such as $r$ and $|Eng|$. That is escaping probability and prey's energy level, respectively. These two variables are essential in the HHO strategy. The condition applied is r < 0.5, which shows the higher possibility for prey escaping. The $r \geq 0.5$ condition indicates chances of running are reduced. The following sections provide brief illustrations about each tactic.

a) Soft besiege

Rabbit is the usual prey of Hawks which has adequate energy level to escape from being attacked that can be illustrated by $|Eng| \geq 0.5$ and $r \geq 0.5$. On the other hand, Hawks executes its tactics to minimize its prey energy level by encircling it by soft besieges. This strategy is done before running the surprise pounce of the Hawk. The mathematical derivation is shown in Equation (14) below [38].

$$\begin{cases} X(t+1) = \Delta X(t) - E|JX_{rabbit} - X(t)| \\ \quad\quad \Delta X(t) = X_{rabbit} - X(t) \\ \quad\quad J = 2(1 - r_5), \quad r_5 \in [0,1] \end{cases} \tag{14}$$

Where:

$J$ => Prey's jump power,

$r_5$ => Random variables,

$X(i + 1)$ => Hawk's position in the next iteration of $i$,

$X(i)$ => Hawk's position vector in the current iteration $i$,

$Eng$ => Energy of prey,

$\Delta X(i)$ => Variations between the prey's position vector and current location in iteration $i$.

b) Hard besiege

Here, the rabbit is got tired due to the soft besiege, and its energy is reduced, where the values of the variables are $|E| < 0.5$ and $\geq 0.5$.

The soft besiege makes the prey tired (energy level reduced), where the values of variables are $|Eng| < 0.5$ and $\geq 0.5$. Then, hawks process the surprise pounce hardly and encircle the prey. The derivation is shown in Equation (15) [38].

$$X(t + 1) = X_{rabbit}(t) - E|\Delta X(t)| \tag{15}$$

c) Soft besiege with rapid dives

In this stage, the prey still has some energy for escape with the variable's value of $|E| \geq 0.5$ and $r < 0.5$. Therefore, the Hawk is smartly encircling the rabbit, and before the performance of surprise pounce, it patiently dives. Here, two steps are utilized to update the hawks' position, and the event of dive is known as intelligent soft besiege. The Hawk moves forward to the rabbit with the help of the prey's next move. Equation (16) [38] is derived from the action above.

$$Y = X_{prey}(i) - Eng|JX_{prey}(i) - X(i)| \tag{16}$$

Then, the possible results are compared based on movements executed at earlier dive to decide whether it is a good dive.

The Hawk generates irregular dives when decisions are not taken based on the levy flight (LF) method. This generation is formulated in Equation (17) [38].

$$Z = Y + S \times LF(Dim) \tag{17}$$

Where:

$Dim$ => Dimension of solutions,

$S$ => Random vector of size $1 \times dim$,

$LF$ => Levy Flight function.

The levy flight function is calculated by using Equation (18) [38].

$$LF(x) = 0.01 \times \frac{u \times \sigma}{|v|^{\frac{1}{\beta}}}, \quad \sigma = \left(\frac{\Gamma(1+\beta) \times \sin(\frac{\pi\beta}{2})}{\Gamma(\frac{1+\beta}{2}) \times \beta \times 2^{(\frac{\beta-1}{2})}}\right)^{\frac{1}{\beta}} \quad (18)$$

Where:

$\beta$ => Default constant which is automatically set as 1.5,

$u, v$ => Random values within [0,1].

Hence, the position of the Harris hawks is updated with the progressive rapid dives calculated using Equation (19) [38].

$$X(t+1) = \begin{cases} Y & if\ F(Y) < F(X(t)) \\ Z & if\ F(Z) < F(X(t)) \end{cases} \quad (19)$$

Where:

$Y\ and\ Z$ => Next location of the new iteration $t$.

This new iteration is performed by utilizing the equations (18) and (19).

d) Hard besiege with progressive rapid dives

Here, the values of variables are $|Eng| < 0.5$ and $r < 0.5$. In this situation, hawks try to reach the rabbit by rapid dives because it has no more energy to escape. This action is performed before the surprise pounce to catch the rabbit. In this Equation, $Z$ is updated by utilizing Equation (20), and $Y$ is accomplished by using Equation (20) [38].

$$Y = X_{prey}(i) - E|JX_{prey}(i) - X_m(i)| \quad (20)$$

The $X_m(i)$ is obtained by utilizing Equation (12), the four strategies are helping hawks catch their prey.

With the help of exploration and exploitation techniques of Harris Hawks Optimization, the weightage of the classification technique's parameters is enhanced in the proposed work.

### 4.3.5. The proposed G-HHO for brain tumor detection

By utilizing the presented enhanced HHO algorithm, the DCNN training process can find the optimum weights for tuning the DCNN classifier. The hyperparameters tuning is done by adopting GWO [39] and HHO. The HHO is improved by utilizing the technique GWO; therefore, the performance of the deep learning technique can be significantly enhanced on performance level. The proposed work has implemented the HHO strategy to select the optimization problem in choosing the hyper-parameters. This algorithm enables a better searching strategy; therefore, the performance and results can be most effective. As discussed in the previous section, DCNN performance is improved by tuning the parameters of CNN with the G-HHO strategy. The parameters utilized for this work are three convolutional Layers, two fully connected layers, and finally, one classification layer. Each convolutional layer is connected by three ReLu layers and a max-pooling layer. The pseudocode of the proposed method is presented in Algorithm (1):

**Algorithm (1): Pseudocode of the Proposed G-HHO**

Inputs: *N* denotes population size, and the maximum number of iterations is i

Outputs: determine the best fitness value, RMSE

Initialize the random population $X_i (i = 1, 2, ..., N)$

While (it is not met the stopping condition) do

    Compute the hawks' fitness values

    Set $X_{prey}$ as the prey's best location

    For (every Hawk ($X_i$)) do

        Update the $Eng_0$ (energy level at initial stage) and $j$ (strength of jumb)

        $Eng_0$=2 rand () -1, j=2(1-rand())

        Update the Eng by utilizing equation (13)

        if ($|Eng| \geq 1$) then                               ► Exploration phase

    vector location updation utilizing equation (10)

        if ($|Eng| < 1$) then                                ► Exploitation phase

            if ($r \geq 0.5$ and $|Eng| \geq 0.5$) then      ► Soft besiege

                vector location updating and utilizing Equation (14)

            else if ($r \geq 0.5$ and $|Eng| < 0.5$) then      ► Hard besiege

                Update the vector location by utilizing Equation (15)

            else if ($r < 0.5$ and $|Eng| \geq 0.5$) then      ► Soft besiege

with progressive quick dives

    vector updating using utilizing Equation (19)

    else if ($r < 0.5$ and $|Eng| < 0.5$) then     ► Hard besiege

with progressive quick dives

    Update the vector location by utilizing Equation (20)

    Return $X_{prey}$ (optimistic result)

then

Initialize new population $W_j$ where J ϵ 1,2,3,.., NP of Grey wolf Optimization

Initialization of a, $\vec{C}$, and $\vec{A}$,

    Compute the fitness value by utilizing every search agent's objective function

Allot

    $W_{alpha}$ →

    the search agent's best fitness value

    $W_{beta}$ →

    the search agent's second-best fitness value

    $W_{delta}$ →

    the search agent's second-best fitness value

    Upgrade the location of $W_j$ (Grey Wolf Optimization) utilizing Harris Hawk

While ( i⟨$i_{max}$) for each search agent

    Upgrade the location of the present search agent

 End for

    Upgrade the value of a, $\vec{C}$, and $\vec{A}$,

    Upgrade the value of $W_{alpha}$, $W_{beta}$ and $W_{delta}$

i = i +1

    end while

Return $W_{alpha}$

## 5. Experimental setup

This section integrated with the proposed brain tumor detection outcome through an experimental analysis-based Deep CNN model. The proposed model is analyzed and validated with the existing techniques. To prove the performance of the proposed DCNN-G-HHO is highly effective than existing. Therefore, the performance analysis parameters such as accuracy, recall, precision, f1-measure are considered. Furthermore, the datasets are initially divided into training and testing, where 70% of datasets are utilized for training, and the remaining 30% are being used for testing. The implementation of the suggested DCNN-G-HHO is done through the Python platform. The Python source code of DCNN-G-HHO can be accessible in [40].

The following sub-sections are explanations on data description, Implementation components, and performance evaluation.

### 5.1. Dataset description

The brain MRI images are obtained from the open-source datasets from the internet Kaggle [41]. In this work, the MRI scanned image-based datasets are downloaded and utilized. The obtained datasets are fed into Deep CNN-IWOA. That mainly focuses on detecting brain tumor detection and classification by using the deep learning technique. Therefore, MRI-based brain tumor classification can be obtained from massive datasets. The total number of input images is utilized before augmentation is 248, and once after augmentation is completed, the input images are turned into 2073. The sample input images of the proposed work are given in Figure (3). Also, Figure (4) shows the brain tumor image.

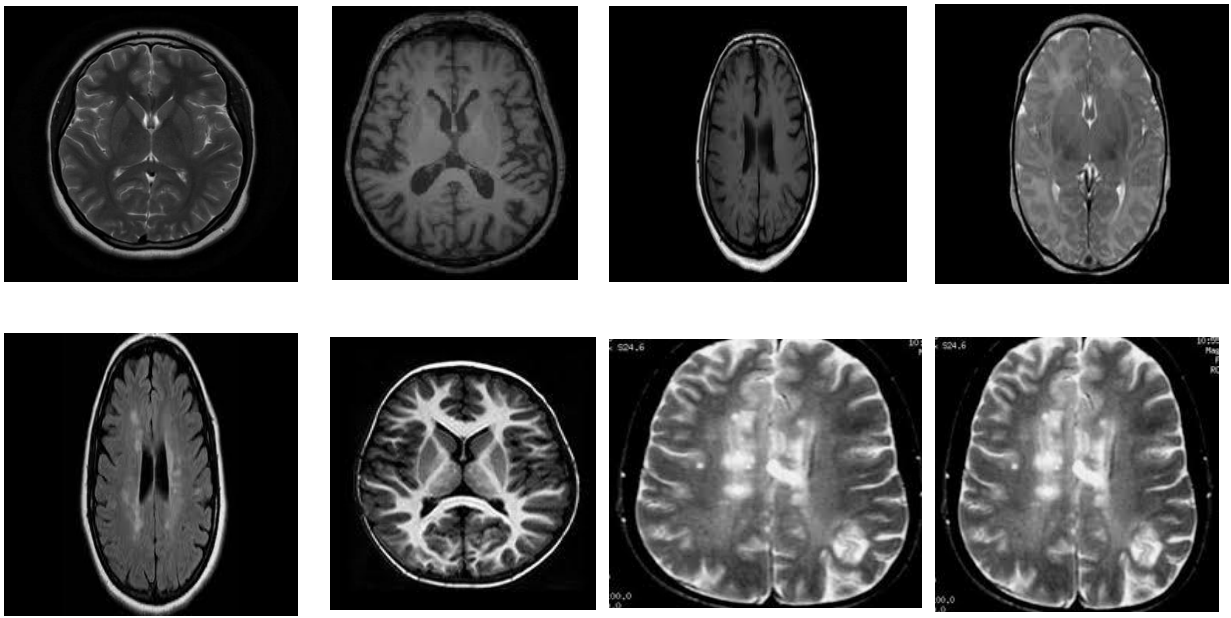

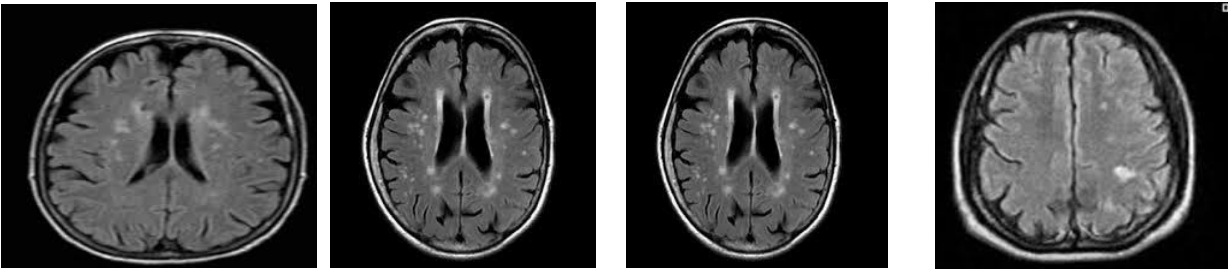

**Figure (3): Sample input images**

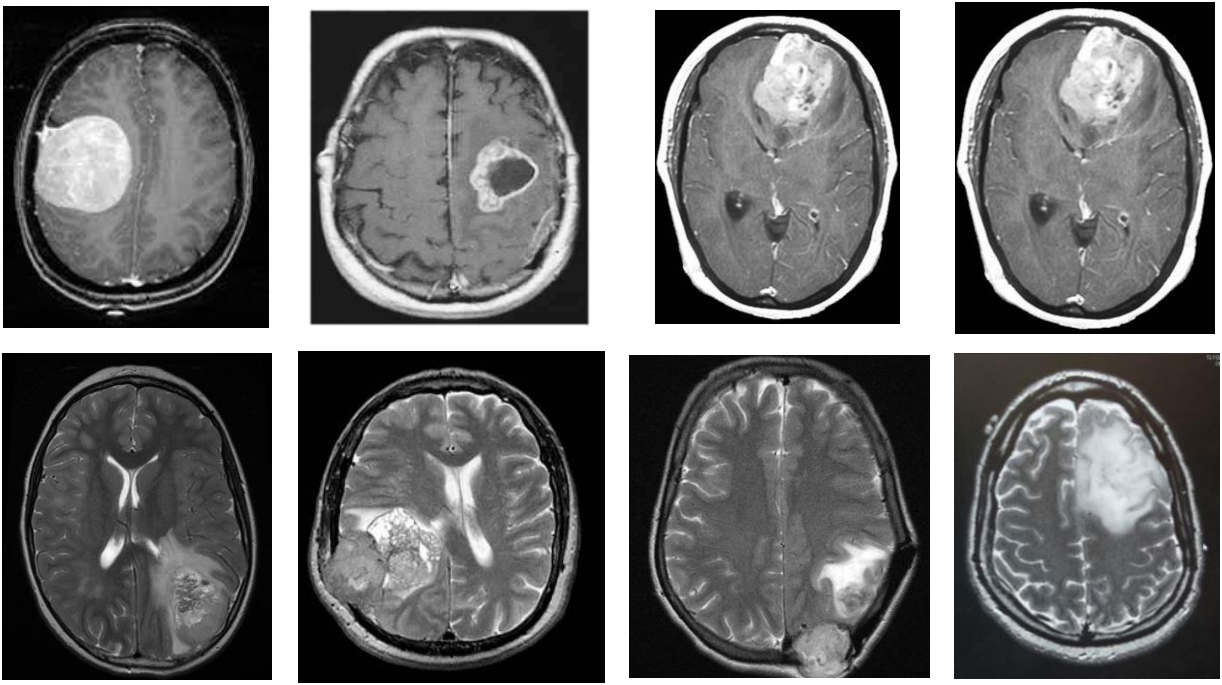

**Figure (4): The brain tumor detected image**

The following Figure (4) shows the sample images of brain tumor affected MRI images. In each MRI brain image, a strange shape contaminated is a brain tumor. Figure (5) shows the augmentation samples of the proposed work.

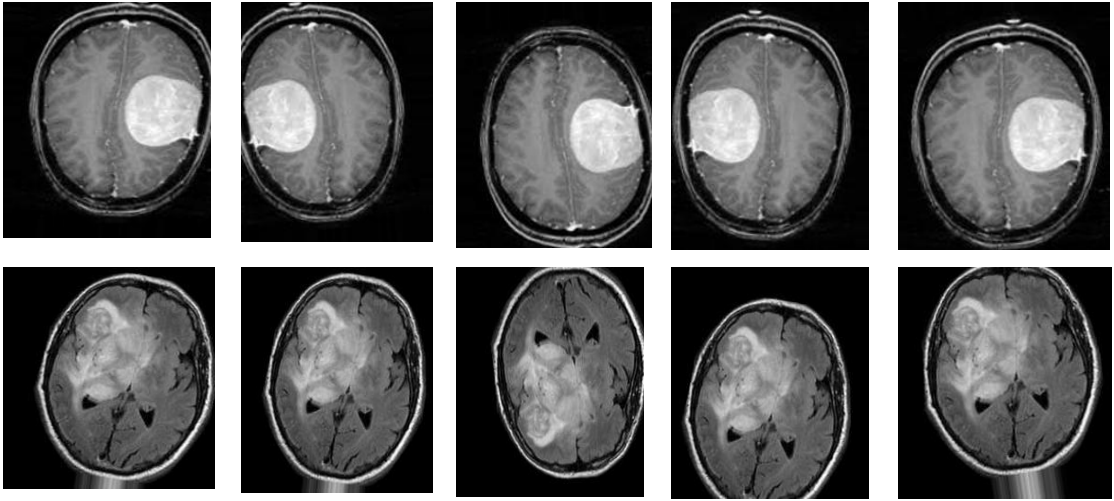

**Figure (5): Augmented images**

The other section describes the implementation requirements of the proposed model.

## 5.2. Implementation components

In this part, the experimental results are discussed and clarified. By using the python software, the implementation process is executed. The performance analysis is conducted to evaluate the proposed approach with the existing precision, F-measure, accuracy, and recall methods. For the training process, 70% of the dataset is used, and for the testing process, 30% of the data set is used for executing the process. The execution process is performed on Windows 10 operating system with 16 GB RAM. The epoch values are taken from 0 to 500. According to [42], the best accuracy is achieved by the 1024 batch size. The author states that based on their results, the higher the batch size, the higher the accuracy, meaning that the batch size has a massive impact on the CNN performance. The batch size is set to 1024 for each epoch on this premise. Also, the kernel size is considered at 5 X 5. Additionally, our proposed classifier is evaluated against the state-of-the-art algorithms regarding execution time and memory usage on the MRI dataset.

## 6. Evaluation and result analysis

This section is divided into three sub-sections: performance evaluation, comparative performance analysis, and statistical performance analysis.

## 6.1. Performance evaluation

The presented approach can be estimated by some statistical measurements, such as true negative (TN), true positive (TP), False Negative (FN), and False positive (FP) [43]. These values are represented as a confusion matrix in Table (3).

Table (3): **Confusion matrix to describe statistical values**

| n = 2065 | Output | | |
|---|---|---|---|
| | Perfect<br>NO | Defected<br>YES | |
| Perfect<br>NO | TP = 1075 | FP = 10 | 1085 |
| Defected<br>YES | FN = 51 | TN = 929 | 980 |
| | 1126 | 939 | |

By using the following equations, these statistical parameter values are evaluated, and those equations are formulated below.

*Accuracy*

Accuracy refers to the ratio of the true patterns to the summation of entire patterns. It can be expressed as

$$Accuracy = \frac{TN + TP}{TN + TP + FN + FP} = \frac{929 + 1075}{dx2065} = 0.97$$

Therefore, the accuracy for the proposed DCNN-G-HHO is 0.97.

*Precision*

Precision can be described as the ratio of true positive to the summation of entire positive patterns. It can be expressed as

$$Precision = \frac{TP}{TP + FP} = \frac{1075}{1075 + 10} = 0.99$$

Thus, precision for the introduced DCNN-G-HHO is 0.99

### *Recall*

This recall parameter is used for calculating the false positive and true positive values, and also it can be called sensitivity. It can be expressed as

$$Recall = \frac{TP}{TP + FN} = \frac{1075}{1075 + 51} = 0.95$$

The recall value for the suggested DCNN-G-HHO is 0.95.

### *F-measure*

To measure the value of F-measure, it requires both precision and recall values. It can be expressed as

$$Fmeasure = 2 \times \frac{Precision \times Recall}{Precision + Recall} = 2 \times \frac{0.99 \times 0.954}{0.99 + 0.954} = 0.97$$

The F-measure value for the proposed approach is 0.97.

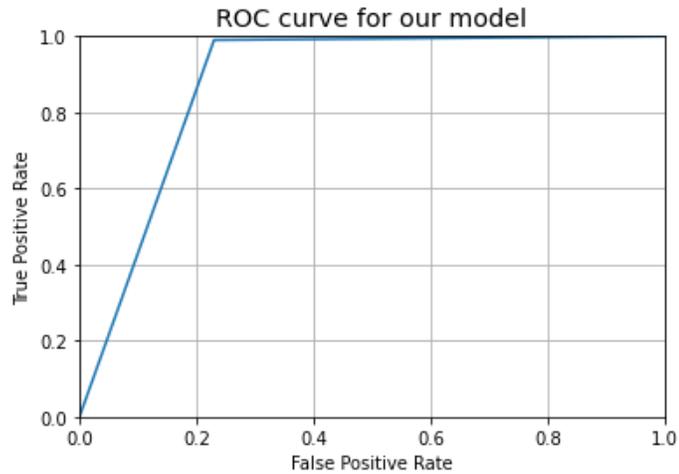

**Figure (6): ROC curve for the proposed model**

Figure (6) shows that the x-axis represents the false positive rate, and the y- axis represents the true positive rate.

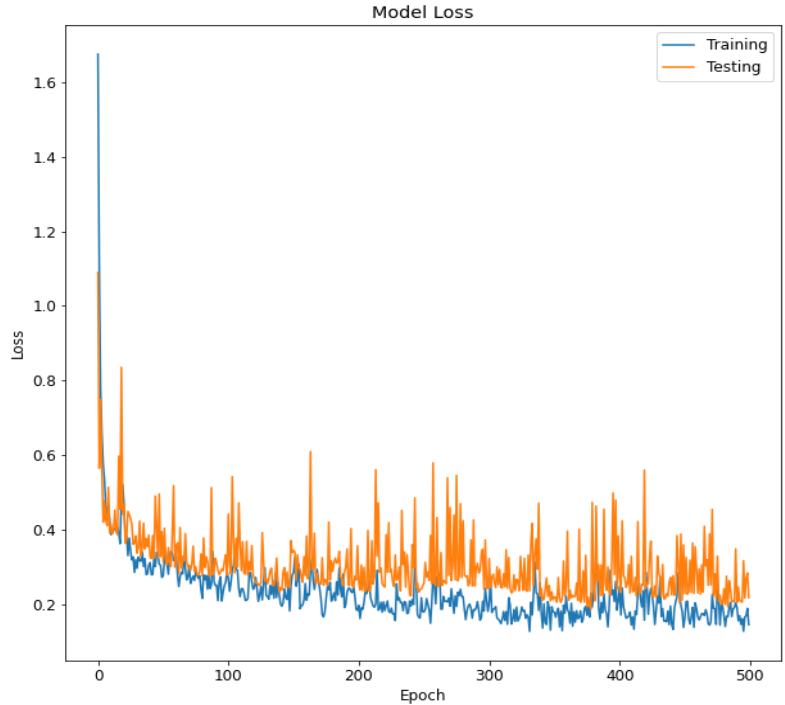

**Figure (7): The graphical illustration of Loss**

Figure (7) shows that the x-axis represents the epoch value, and the y-axis represents the loss value. The blue color represents the training process, and the orange color represents the testing process. The epoch values are taken from 0 to 500.

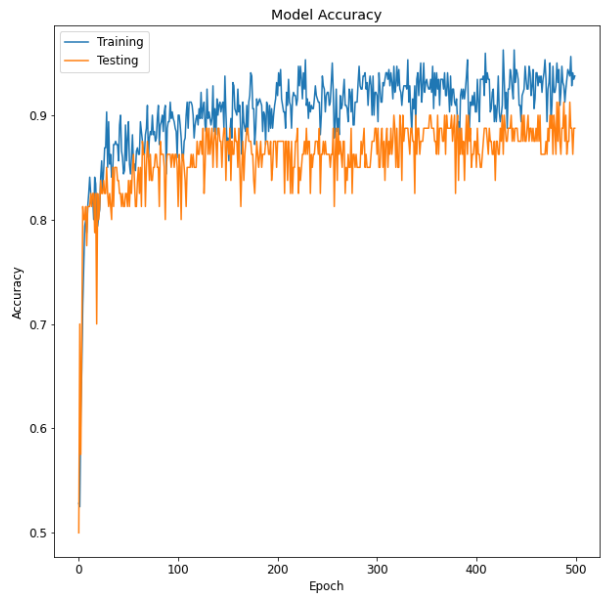

**Figure (8): Accuracy for the proposed model**

In Figure (8), the x-axis represents the epoch, and the y-axis represents the accuracy. The epoch values are taken from 0 to 500. The blue color represents the training process, and the orange color represents the testing process.

## 6.2. Comparative performance analysis

In this part, the comparative analysis is described for the presented approach and the existing approaches such, as (Long Short-Term Memory (LSTM), Artificial Neural Network (ANN), CNN, SVM, Wide & Deep Learning, Deep Factorization Machines (DeepFM), SR-FCM-CNN, and Whale Harris Hawks optimization with DeepCNN (WHHO-based DeepCNN)). The proposed and current methods values are given in Table (4).

Table (4): Values for the comparative analysis

| Classification technique | Accuracy | precision | Recall | F-measure |
| --- | --- | --- | --- | --- |
| LSTM | 0.91 | 0.93 | 0.88 | 0.92 |
| ANN | 0.87 | 0.86 | 0.85 | 0.86 |
| CNN | 0.94 | 0.94 | 0.89 | 0.93 |
| SVM | 0.90 | 0.89 | 0.86 | 0.88 |
| Wide & Deep Learning | 0.93 | 0.90 | 0.87 | 0.92 |
| DeepFM | 0.88 | 0.92 | 0.87 | 0.89 |
| SR-FCM-CNN | 0.89 | 0.95 | 0.90 | 0.94 |
| WHHO-based DeepCNN | 0.96 | 0.94 | 0.91 | 0.93 |
| The proposed DCNN-G-HHO | 0.97 | 0.99 | 0.95 | 0.97 |

The above table denotes the results of the existing and presented approach. The accuracy obtained for the brain tumor detection from our approach is 0.97. The accuracy of the competitive algorithms, such as LSTM, ANN, CNN, SVM, Wide & Deep Learning, DeepFM, SR-FCM-CNN, and WHHO-based DeepCNN are 0.91, 0.87, 0.94, 0.90, 0.93, 0.88, 0.89, and 0.96, respectively. The precision level for the presented approach is 0.99, where for the existing techniques, LSTM, ANN, CNN, SVM, Wide & Deep Learning, DeepFM, SR-FCM-CNN, and WHHO-based DeepCNN are 0.93, 0.86, 0.94, 0.89, 0.90, 0.92, 0.95, and 0.94, respectively. This shows that the precision level for the presented approach was high and performed well with an effective result. The recall value for the suggested method is 0.94, where for the existing algorithms, LSTM, ANN, CNN, SVM, Wide & Deep Learning, DeepFM, SR-FCM-CNN, and WHHO-based DeepCNN are 0.88, 0.85, 0.89, 0.86, 0.87, 0.87, 0.90, and 0.91, respectively. At last, the F-measure value for the presented method is 0.96, where for the existing methods, LSTM, ANN, CNN, SVM, Wide & Deep Learning, DeepFM, SR-FCM-CNN, and WHHO-based DeepCNN are 0.92, 0.86, 0.93, 0.88,

0.92, 0.89, 0.94, and 0.93, respectively. Therefore, the presented approach is more efficient in accuracy, precision, recall, and F-measure from this comparative analysis than the existing approaches. The comparative analysis graph is represented in the bar diagram shown in Figure (9).

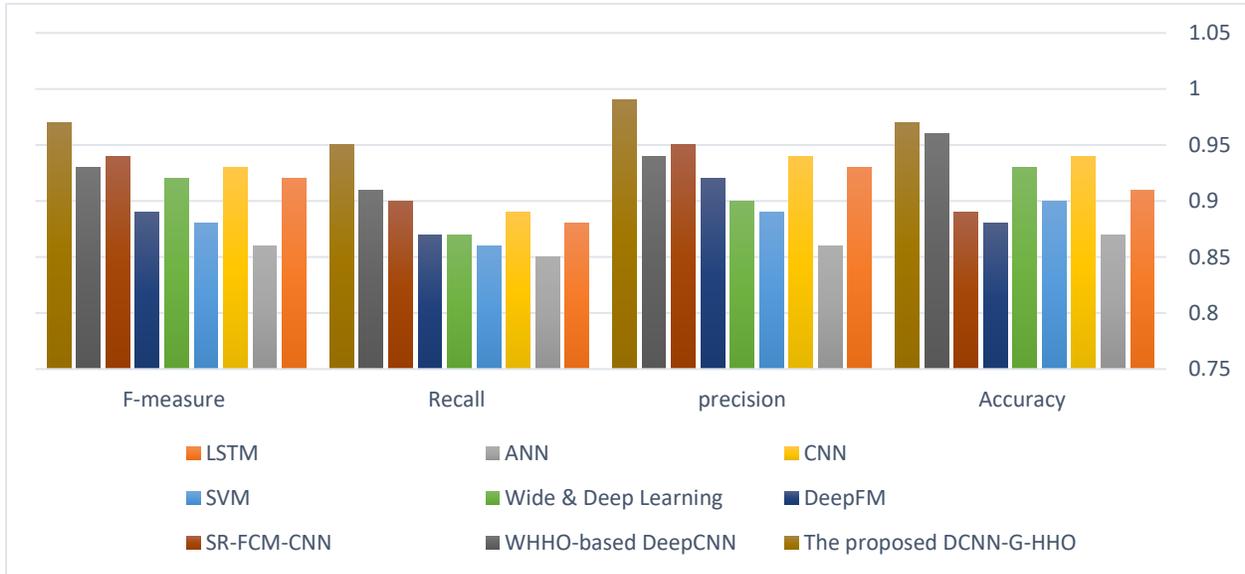

**Figure (9): Bar diagram for the comparative analysis**

The x-axis represents brain tumor detection performance metrics in the above Figure, such as accuracy, precision, recall, and F-measure. The y-axis represents the parameter values. The presented approach is denoted by green color. LSTM, ANN, CNN, SVM, Wide & Deep Learning, DeepFM, SR-FCM-CNN, and WHHO-based DeepCNN, are indicated by golden, dark gray, brown, dark blue, green, light blue, yellow, light gray, and orange. This coloring clearly shows that the presented approach is performed efficiently in accuracy, precision, recall, and f-measure compared with the existing algorithms. Thus, the proposed method is most preferred for detecting brain tumors.

### 6.3. Statistical performance analysis

This section analyses the overall performance benchmarking of the algorithms (execution time/memory consumption) based on the augmented MRI image dataset. Table (2) presents the execution time with memory consumption for the 30 solutions obtained by the proposed DCNN-G-HHO and its nine state-of-the-art algorithms. We observe that our suggested model exhibited a shorter execution time for detecting and classifying the brain tumors in the MR images. Similarly, our proposed method consumed less memory than its counterpart techniques. Surprisingly, KNN had a slower execution time than the other techniques, whereas SVM required the highest memory allocation than its competitive algorithms. Generally, the suggested DCNN-G-HHO technique had a faster execution and consumed less memory than the other classifiers.

Table (2): Average execution time with memory consumption for the 30 solutions obtained by the proposed DCNN-G-HHO against its counterpart algorithms

| Algorithms | Execution time | Memory consumption (MB) |
|---|---|---|
| **LSTM** | 26.122 | 105.109 |
| **ANN** | 37.128 | 102.085 |
| **KNN** | 41.013 | 95.101 |
| **Deep KNN** | 38.891 | 97.982 |
| **SVM** | 40.172 | 103.138 |
| **Wide & Deep Learning** | 39.183 | 101.283 |
| **DeepFM** | 30.837 | 98.012 |
| **SR-FCM-CNN** | 22.193 | 79.373 |
| **WHHO-based DeepCNN** | 23.148 | 89.172 |
| **The proposed DCNN-G-HHO** | 18.091 | 74.183 |

## 7. Conclusion

The main intention of the proposed DCNN-G-HHO was to provide an automated detection model for brain tumors. Therefore, a deep learning-based CNN was considered and enhanced by adopting a hybrid optimization algorithm with a combination of GWO and HHO techniques. Moreover, segmentation plays a significant role in detecting tumors. Accordingly, an Otsu thresholding technique is utilized. As a result, effective segmentation and classification were achieved better results.

The technique's performance was ensured by comparing it with the existing technique on huge augmented MRI images in terms of accuracy, precision, recall, f-measure, execution time, and memory consumption. At first glance, the performance comparison proved that the suggested DCNN-G-HHO was highly effective than current techniques. With 0.97 accuracies for brain tumor diagnosis, the recommended DCNN-G-HHO method outperformed current approaches. This demonstrates the approach's accuracy and effectiveness. The proposed approach has a recall of

0.95, whereas the existing methods had lower recollections than our method. Finally, the F-measure for the proposed technique is 0.96. Therefore, the proposed methodology is more efficient than the current methods regarding accuracy, precision, recall, and F-measure. Additionally, the overall performance of our proposed approach was benchmarked in terms of execution time and memory use on the enhanced MRI image dataset. The suggested DCNN-G-HHO was faster at identifying and categorizing brain tumor cancers on the MR images. Similarly, our solution used less RAM than its competing methods. Surprisingly, KNN took longer than the other methods, but SVM consumed the most memory. The DCNN-G-HHO approach was quicker and used less memory than the different classifiers.

The main contributions of the newly introduced DCNN-G-HHO were:

- High accuracy and precision of detecting and classifying brain tumors due to the use of G-HHO and Otsu thresholding combined with DCNN;
- Massive augmented MRI image datasets were used to provide reliable results;
- The performance of the suggested DCNN-G-HHO is improved and enabled better detection and classification results.

Thus, we have designed DCNN-G-HHO as a combination of DCNN and G-HHO with few exploratory and exploitative mechanisms. It is also viable to use other evolutionary strategies such as mutation and crossover, multi-swarm and multi-leader structures, evolutionary updating structures, and chaos-based phases. These operators and concepts are advantageous for future works. Binary and multi-objective variants of G-HHO may be developed in future developments. Additionally, it may be used to address a variety of issues in engineering and other industries. Another intriguing avenue is to examine various ways of coping with constraints in real-world restricted scenarios.

Because of its high accuracy, DCNN-G-HHO could be deployed on different applications. Besides, the modifications, as well as hybridizations of the algorithm, have been a hot spot for scholars. But still, there is room for others to work in improving our introduced algorithm as it can be hybridized with most recent swarm algorithms, such as backtracking search optimization algorithm [44, 45], the variants of evolutionary clustering algorithm star [46–49], chaotic sine cosine firefly algorithm [50], and hybrid artificial intelligence algorithms [51]. Furthermore, DCNN-G-HHO can be applied to more complex and real-world applications to explore more deeply the advantages and drawbacks of the algorithm or improve its efficiencies, such as engineering application problems [50], laboratory management [52], e-organization and e-government services [53], online analytical processing [54], web science [55], the Semantic Web ontology learning [56], cloud computing paradigms [57][58][59][60], and evolutionary machine learning techniques [49, 61, 62].

## Acknowledgments

The authors would like to express their sincere gratitude to their respective institutions for providing the facilities and ongoing assistance necessary to perform this research.

## Compliance with ethical standards

**Conflict of interest:** There is no conflict of interest.

**Funding:** This research did not receive any specific grant from funding agencies in the public, commercial, or not-for-profit sectors.

from Corpora Using Adaptive Evolutionary Clustering Algorithm. Complex Intell Syst